\documentclass[aps,prb,twocolumn,showpacs,preprintnumbers,amsmath,amssymb,superscriptaddress]{revtex4}

\usepackage{graphicx}
\usepackage{dcolumn}
\usepackage{bm}
\usepackage{color}

\newcommand{\ket}[1]{|#1\rangle}
\newcommand{\e}{\varepsilon}
\newcommand{\s}{\sigma}
\newcommand{\up}{\uparrow}
\newcommand{\down}{\downarrow}
\newcommand{\expect}[1]{\langle #1 \rangle}

\begin{document}

\title{Underscreened Kondo effect in quantum dots coupled to ferromagnetic leads}

\author{Ireneusz Weymann}
\email{weymann@amu.edu.pl} \affiliation{Department of Physics,
Adam Mickiewicz University, 61-614 Pozna\'n, Poland}
\affiliation{Physics Department, Arnold Sommerfeld
Center for Theoretical Physics and Center for NanoScience, \\
Ludwig-Maximilians-Universit\"at, Theresienstrasse 37, D-80333
Munich, Germany}
\author{L\'{a}szl\'{o} Borda}
\affiliation{Department of Theoretical Physics and Research Group
  ``Physics of Condensed Matter'' of the Hungarian Academy of
  Sciences,\\
Budapest University of Technology and Economics, Budafoki \'{u}t 8.,
H1111 Budapest, Hungary}
\affiliation{Physikalisches Institut and Bethe Center for
Theoretical Physics, Universit\"at Bonn, Nussallee 12, D-53115
Bonn, Germany}

\date{\today}

\begin{abstract}
We analyze the equilibrium transport properties of underscreened
Kondo effect in the case of a two-level quantum dot coupled to
ferromagnetic leads. Using the numerical renormalization group
(NRG) method, we have determined the gate voltage dependence of
the dot's spin and level-resolved spectral functions. We have
shown that the polarization of the dot is very susceptible to spin
imbalance in the leads and changes sign in the middle of the $S=1$
Coulomb valley. Furthermore, we have also found that by
fine-tuning an external magnetic field one can compensate for the
presence of ferromagnetic leads and restore the Kondo effect in
the case of $S=\frac{1}{2}$ Coulomb valley. However, the
underscreened Kondo effect cannot be fully recovered due to its
extreme sensitivity with respect to the magnetic field.
\end{abstract}


\maketitle

\section{Introduction}

Ever since the observation of the resistivity anomaly in normal
metals at low temperatures,~\cite{HaasPhysica36} the Kondo
problem~\cite{kondo64} has been investigated perpetually both
experimentally and theoretically.~\cite{hewson_book,
goldhaber-gordon_98,cronenwett_98} As a result of that intensive
research, in many respects, the Kondo effect is well understood
now. Very recently the interplay between the Kondo effect and
other many-body phenomena, such as superconductivity or
ferromagnetism, has attracted a lot of interest. That attention
was mainly motivated by the recent advances in nanofabrication
which have opened the possibility to attach
superconducting~\cite{SchoenenbergerPRL02,TaruchaPRL07} or
ferromagnetic~\cite{pasupathy_04,heerschePRL06,
hamayaAPL07,hamayaPRB08,hauptmannNatPhys08,parkinNL08} leads to
molecules or semiconducting quantum dots. In the following we will
focus our attention on the case of ferromagnetic leads coupled to
a quantum dot exhibiting the Kondo effect.

From the theoretical side, a consensus was found that single level
quantum dots (i.e. relatively small dots with level spacing
$\delta$ larger than the hybridization $\Gamma$) attached to
ferromagnetic leads exhibit finite spin asymmetry as a result of
the spin imbalance in the leads, which gives rise to a splitting
and suppression of the Kondo resonance.~\cite{LopezPRL03,
martinekPRL03_2,martinekPRL03,ChoiPRL04} This spin asymmetry can,
however, be compensated by means of an external magnetic field or
by tuning the gate voltage
properly.~\cite{martinek_PRB05,sindelPRB07} In that way the strong
coupling Kondo fixed point can be reached, with a fully developed
Kondo resonance and a somewhat reduced Kondo temperature. This
expectation has recently been confirmed
experimentally.~\cite{pasupathy_04,heerschePRL06,
hamayaAPL07,hamayaPRB08,hauptmannNatPhys08,parkinNL08}

While in many cases the quantum dots can successfully be modelled
by only a single local level, there are few exceptions when the
finite level spacing affects the low energy physics of such a
device considerably. An isolated quantum dot with even number of
electrons can have spin triplet $S=1$ ground state, if the level
spacing between the two single particle levels closest to the
Fermi energy is anomalously small, i.e. smaller than the exchange
coupling between the two electrons situated on those levels. If
such a device is attached to only one lead then the system will
exhibit the so-called {\em underscreened} Kondo
effect.~\cite{NozieresJP80,
LeHurPRB97,ColemanPRB03,PosazhennikovaPRL05,MehtaPRB05} A single
lead can only screen half of the local spin below the Kondo
temperature $T_K$ while the residual spin-a-half object is left
unscreened with a ferromagnetic Kondo coupling to the rest of the
conduction electrons. Such a ferromagnetic Kondo coupling is known
to be irrelevant in renormalization group sense, as it scales to
zero when the energy scale is lowered, resulting in a Fermi liquid
plus a decoupled $S=\frac{1}{2}$ object at zero temperature. This
behavior is reached in a non-analytic fashion resulting in
singular terms in e.g. the single particle scattering rate. Such
kind of behavior was recently termed as {\em singular Fermi
liquid}.~\cite{ColemanPRB03,MehtaPRB05}

In a real quantum dot subject to a transport experiment the
situation is slightly more complex as the dot has to be coupled to
two leads to drive current through it. Even if the second lead is
just a weakly coupled probe, its presence introduces a second
energy scale, $T_K'\ll T_K$, at which the residual spin-a-half is
screened by the other lead mode. The screening takes place in two
stages and the separation between the energy scales $T_K$ and
$T_K'$ can be tuned by the asymmetry between the coupling to the
leads. Since the Kondo temperature is exponentially sensitive to
the coupling, a realistic value of asymmetry is enough to separate
the two stages completely. In such a case, if the other
experimentally relevant energy scales, such as temperature $T$ or
magnetic field $B$, lie in between the two Kondo temperatures
$T_K'\ll B,T < T_K$, then the system is well described by the
underscreened Kondo model. In fact, very recently, underscreened
Kondo effect was observed experimentally in molecular quantum dots
coupled to nonmagnetic leads.~\cite{Roch09}

In the present paper we focus our interest on the interplay of
underscreened Kondo effect and itinerant electron ferromagnetism
in the leads. To that end we consider a two level quantum dot
coupled to a single reservoir of conduction electrons exhibiting
spin imbalance. We assume that the second Kondo temperature is
always much smaller than the experimental temperature, therefore
we consider the second lead as a weakly coupled probe only.

\section{Theoretical description}

\begin{figure}[t]
  \includegraphics[width=0.5\columnwidth,clip]{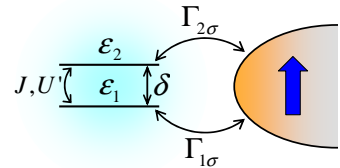}
  \caption{\label{Fig:1}
  (color online) Schematic of a two-level quantum dot coupled to
  a ferromagnetic reservoir with spin polarization $p$. The dot level
  energies are denoted by $\e_1$ and $\e_2$, $\delta$ is the level spacing,
  while $U'$ and $J$ describe the inter-level Coulomb correlations and
  spin exchange interaction. The $j$th dot level is coupled
  to ferromagnetic electrode with strength $\Gamma_{j\s}$.}
\end{figure}

The considered system consists of a two-level quantum dot coupled
to a ferromagnetic reservoir, see Fig.~\ref{Fig:1}, and its
Hamiltonian is given by
\begin{equation}\label{Eq:H}
H = H_{\rm FM} + H_{\rm QD} + H_{\rm tun},
\end{equation}
where the first term describes the noninteracting itinerant
electrons in ferromagnetic lead, $H_{\rm FM} = \sum_{k\s} \e_{k\s}
c^\dag_{k\s} c_{k\s}$, where $c^\dag_{k\s}$ is the electron
creation operator with wave number $k$ and spin $\s$, while
$\e_{k\s}$ is the corresponding energy. The second part of the
Hamiltonian describes a two-level quantum dot and is given by
\begin{eqnarray}\label{Eq:HQD}
H_{\rm QD} &=& \sum_{j\s} \e_{j} d^\dag_{j\s} d_{j\s} + U
\sum_j n_{j\uparrow} n_{j\downarrow}\nonumber\\
&& + U'\sum_{\s\s'} n_{1\s}n_{2\s'} + JS^2 + B S^z,
\end{eqnarray}
where $n_{j\s} = d^\dag_{j\s} d_{j\s}$ and $d^\dag_{j\s}$ creates
a spin-$\s$ electron in the $j$th level ($j=1,2$), $\e_{j}$
denotes the corresponding energy of an electron in the dot. Here,
$\e_1 = \e - \delta/2$ and $\e_2 = \e  + \delta/2$, with
$\delta=\e_2-\e_1$ being the level spacing. The on-level
(inter-level) Coulomb correlations are denoted by $U$ ($U'$),
respectively. $J$ is the ferromagnetic exchange coupling ($J<0$)
with $\vec{S}=\vec{S}_1+\vec{S}_2$, where $\vec{S}_j =
\frac{1}{2}\sum_{\s\s'}d^\dag_{j\s}\vec{\s}_{\s\s'}d_{j\s'}$ is
the spin operator for electrons in the dot level $j$ and
$\vec{\s}$ denotes the vector of Pauli spin matrices. The last
term of $H_{\rm QD}$ describes the Zeeman splitting with $B$
($g\mu_B\equiv 1$) being the external magnetic field applied along
the $z$th direction.

Finally, the tunnel Hamiltonian is given by
\begin{equation}
H_{\rm tun} = \sum_{kj\s} t_{j\s} \left( d_{j\s}^\dag c_{k\s} +
c_{k\s}^\dag d_{j\s} \right),
\end{equation}
where $t_{j\s}$ describes the spin-dependent hopping matrix
elements between the $j$th dot level and ferromagnetic lead. The
strength of the coupling between the dot and lead can be expressed
as $\Gamma_{j\s} = \pi \rho |t_{j\s}|^2$, where $\rho$ is the
density of states (DOS) in the lead. In the following, we assume a
flat band of width $2D$ and use $D\equiv 1$ as energy unit, as not
stated otherwise. Note that by assuming constant DOS the whole
spin-dependence has been shifted into the coupling constants.
While this assumption may simplify the calculations, it does not
affect the low-energy physics we are interested
in.~\cite{martinekPRL03,ChoiPRL04} To parameterize the couplings,
it is convenient to introduce the spin polarization of
ferromagnetic lead, $p$, defined as $p = (\Gamma_{\uparrow} -
\Gamma_{\downarrow}) / (\Gamma_{\uparrow} + \Gamma_{\downarrow})$,
where we have assumed that each dot level is coupled with the same
strength to the lead, $\Gamma_{j\s} = \Gamma_\s$. Then, the
coupling for the spin-$\up$ (spin-$\down$) electrons can be
written as $\Gamma_{\up(\down)} = (1\pm p)\Gamma$, where $\Gamma =
(\Gamma_\up + \Gamma_\down)/2$. For $p=0$, the couplings do not
depend on spin and the system behaves as if coupled to nonmagnetic
lead. However, for finite spin polarization, $p\neq 0$, the
couplings are spin-dependent, giving rise to an effective exchange
field which may spin-split the levels in the dot, suppressing the
Kondo resonance. Such behavior has been extensively studied
theoretically in the case of single-level quantum
dots,~\cite{LopezPRL03,martinekPRL03_2,martinekPRL03,
ChoiPRL04,martinek_PRB05,sindelPRB07,
UtsumiPRB05,SwirkowiczPRB06,MatsubayashiPRB07,SimonPRB07} where
the usual $S=\frac{1}{2}$ Kondo effect develops, while the effect
of ferromagnetism on the other types of Kondo effect remains to a
large extent unexplored.

In this paper we thus consider the zero-temperature equilibrium
transport properties of two-level quantum dots coupled to
ferromagnetic lead in the case of underscreened Kondo effect. In
order to perform this analysis in most accurate and exact way, we
employ the numerical renormalization group (NRG) method.
\cite{WilsonRMP75,BullaRMP08} The NRG consists in a logarithmic
discretization of the conduction band and mapping of the system
onto a semi-infinite chain with the quantum dot sitting at the end
of the chain. By diagonalizing the Hamiltonian at consecutive
sites of the chain and storing the eigenvalues and eigenvectors of
the system, one can accurately calculate the static and dynamic
quantities of the system. In particular, numerical results
presented here were obtained using the flexible density-matrix
numerical renormalization group (DM-NRG) code, which allows to use
arbitrary number of Abelian and non-Abelian
symmetries.~\cite{Toth_PRB08,FlexibleDMNRG,BudapestNRG} In fact,
exploiting symmetries as much as possible is crucial in obtaining
highly accurate data. Because in the case of ferromagnetic leads
the full spin rotational invariance is generally broken, in the
case of finite spin polarization we have used the Abelian
symmetries for the total charge and spin, while in the case of
$p=0$ we have exploited the full spin $SU(2)$ symmetry.

Using the NRG, we have calculated the expectation value of the
dot's spin as well as the spectral function of the dot,
$A_{jj'\s}(\omega) = -\frac{1}{2\pi}{\rm
Im}[G^R_{jj'\s}(\omega)+G^R_{j'j\s}(\omega)]$, where
$G_{jj'\s}^R(\omega)$ denotes the Fourier transform of the dot
retarded Green's function, $G_{jj'\s}^R(t)= -i\Theta(t)
\expect{\{d_{j\s}(t), d_{j'\s}^\dag(0)\}}$. The diagonal elements
of the spectral function $A_{j\s}(\omega) \equiv A_{jj\s}(\omega)$
are related to the spin-resolved density of states, whereas the
off-diagonal elements $A_{jj'\s}(\omega)$ may be associated with
processes of injecting and removing an electron at different
sites. The symmetrized spin-resolved spectral function of the dot
can be found from
\begin{equation}
  A_\s(\omega) = \sum_{jj'}A_{jj'\s}(\omega).
\end{equation}
On the other hand, the normalized full spectral function is given
by $\pi\sum_\s \Gamma_\s A_\s(\omega)$ and is directly related to
the conductivity of the dot.

\section{Numerical results}

\begin{table}[t]
\caption{\label{tab:energies} The respective eigenstates
$\ket{Q,S^z}$ and eigenvalues $E_{Q,S^z}$ of the decoupled quantum
dot Hamiltonian. Here, $Q= (\sum_{j\s} n_{j\s} - 2)$ and $S^z$ are
the charge and $z$th component of the spin in the dot, while
$\ket{\chi_1\chi_2}$ denotes the local states with
$\chi_j=0,\up,\down,{\rm d}$ for zero, spin-up, spin-down and two
electrons in the level $j$.} \centering
\begin{tabular}{lll}
$n$\;\;&\;\;\; $\ket{Q,S^z}$ & \;\;\;\; $E_{Q,S^z}$ \\\hline
1 & $\ket{-2,0}=\ket{00}$ & $0$\\
2 & $\ket{-1,-\frac{1}{2}}=\ket{\down 0}$ &
$\e-\frac{\delta}{2}+\frac{3J}{4}-\frac{B}{2}$\\
3 & $\ket{-1,-\frac{1}{2}}=\ket{0\down}$ &
$\e+\frac{\delta}{2}+\frac{3J}{4}-\frac{B}{2}$\\
4 & $\ket{-1,\frac{1}{2}}=\ket{\up 0}$ &
$\e-\frac{\delta}{2}+\frac{3J}{4}+\frac{B}{2}$\\
5 & $\ket{-1,\frac{1}{2}}=\ket{0\up}$ &
$\e+\frac{\delta}{2}+\frac{3J}{4}+\frac{B}{2}$\\
6 & $\ket{0,0}=\ket{{\rm d} 0}$ &
$2\e-\delta+U$\\
7 & $\ket{0,0}=\ket{0{\rm d}}$ &
$2\e+\delta+U$\\
8 & $\ket{0,0}=\frac{1}{\sqrt{2}}(\ket{\up\down}-\ket{\down\up})$
&
$2\e+U'$\\
9 & $\ket{0,-1}=\ket{\down\down}$ &
$2\e+U'+2J-B$\\
10 & $\ket{0,0}=\frac{1}{\sqrt{2}}(\ket{\up\down}+\ket{\down\up})$
&
$2\e+U'+2J$\\
11 & $\ket{0,1}=\ket{\up\up}$ &
$2\e+U'+2J+B$\\
12 & $\ket{1,-\frac{1}{2}}=\ket{{\rm d} \down}$ &
$3\e-\frac{\delta}{2}+\frac{3J}{4}+U+2U'-\frac{B}{2}$\\
13 & $\ket{1,-\frac{1}{2}}=\ket{\down{\rm d}}$ &
$3\e+\frac{\delta}{2}+\frac{3J}{4}+U+2U'-\frac{B}{2}$\\
14 & $\ket{1,\frac{1}{2}}=\ket{{\rm d} \up}$ &
$3\e-\frac{\delta}{2}+\frac{3J}{4}+U+2U'+\frac{B}{2}$\\
15 & $\ket{1,\frac{1}{2}}=\ket{\up{\rm d}}$ &
$3\e+\frac{\delta}{2}+\frac{3J}{4}+U+2U'+\frac{B}{2}$\\
16 & $\ket{2,0}=\ket{{\rm dd}}$ & $4\e+2U+4U'$\\
\end{tabular}
\end{table}

Before presenting numerical results, it is instructive to analyze
the eigen-spectrum of the decoupled quantum dot Hamiltonian,
Eq.~(\ref{Eq:HQD}). The respective eigenstates $\ket{Q,S^z}$ and
eigenvalues $E_{Q,S^z}$ of $H_{\rm QD}$ are listed in Table I,
where $Q = (\sum_{j\s} n_{j\s} - 2)$ denotes the charge in the dot
and $S^z$ is the $z$th component of the dot spin. By tuning $\e$
the ground state of the dot changes. In particular, in the absence
of magnetic field and for ferromagnetic exchange coupling ($J<0$),
when $\e > \e_{-2,-1} = \frac{\delta}{2} - \frac{3J}{4}$, the dot
is empty, for $\e =\e_{-1,0} = -\frac{\delta}{2} - U' -
\frac{5J}{4}$, the doublet $\ket{Q=-1,S=\frac{1}{2}}$ and triplet
$\ket{Q=0,S=1}$ states become degenerate, where now $S$ is the
full dot's spin, whereas for $\e =\e_{0,1} = \frac{\delta}{2} - U'
-U + \frac{5J}{4}$, the states $\ket{Q=0,S=1}$ and
$\ket{Q=1,S=\frac{1}{2}}$ are degenerate. Consequently, for
$\e_{-2,-1} > \e > \e_{-1,0}$, the ground state of the decoupled
dot is the doublet and the system will exhibit usual
$S=\frac{1}{2}$ Kondo effect, whereas for $\e_{-1,0}
> \e > \e_{0,1}$ the dot is in the triplet state and the system
will show an underscreened Kondo effect. In calculations we have
taken the following parameters: $\delta = 0.1$, $J=-0.2$, $U=U' =
0.4$, and $\Gamma = 0.04$ (in units of $D=1$). One then finds,
$\e_{-2,-1} = 0.2$, $\e_{-1,0} = -0.2$ and $\e_{0,1}=1$.

\begin{figure}[t]
  \includegraphics[width=0.8\columnwidth,clip]{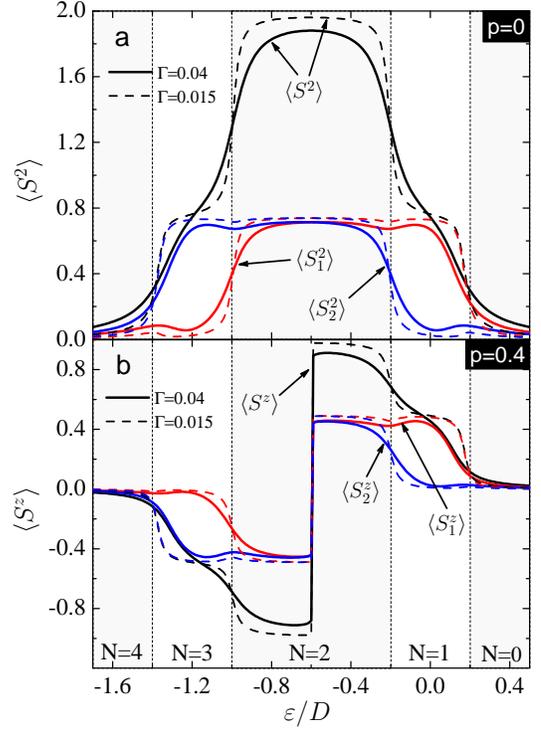}
  \caption{\label{Fig:2}
  (color online) The expectation value of the dot spin operator
  $S^2$ in presence of nonmagnetic (a) and the expectation value of
  dot spin operator $S^z$ for ferromagnetic (b) lead as a function
  of the average level position $\e = (\e_1+\e_2)/2$.
  The solid line corresponds to $\Gamma = 0.04$,
  while the dashed line corresponds to $\Gamma = 0.015$ (in units of $D=1$).
  The expectation values separately for the two dot levels are also shown.
  The parameters are $\delta = 0.1$, $J=-0.2$, $U=U' = 0.4$,
  $B=0$, and in the ferromagnetic case $p=0.4$.
  The shadowed regions indicate Coulomb valleys with different
  electron number in the dot, where $N = \sum_{j\s} n_{j\s}$.}
\end{figure}

We have calculated the expectation value of the dot's spin
operators $S^2$ and $S^z$ in case of nonmagnetic and ferromagnetic
leads, respectively. The results are summarized in
Fig.~\ref{Fig:2}. As the level position is lowered by sweeping the
gate voltage, the dot is tuned through the Coulomb blockade
valleys with electron number $N=1,2,3$ and spin states
$S=1/2,1,1/2$, respectively. When the dot is attached to a
ferromagnetic lead, the spin asymmetry in the hybridization
results in the spin splitting of the dot level, thus leading to a
spin polarized ground state. Precisely in the middle of the $N=2$
Coulomb blockade valley, i.e. for $\e = (\e_{-1,0}+\e_{0,1})/2 =
-U'-\frac{U}{2} = -0.6$, the spin polarization changes sign, as to
the right (left) hand side from that point the electron (hole)
like virtual processes dominate in the renormalization of the
level position.

\begin{figure}[t!]
  \includegraphics[width=0.7\columnwidth,clip]{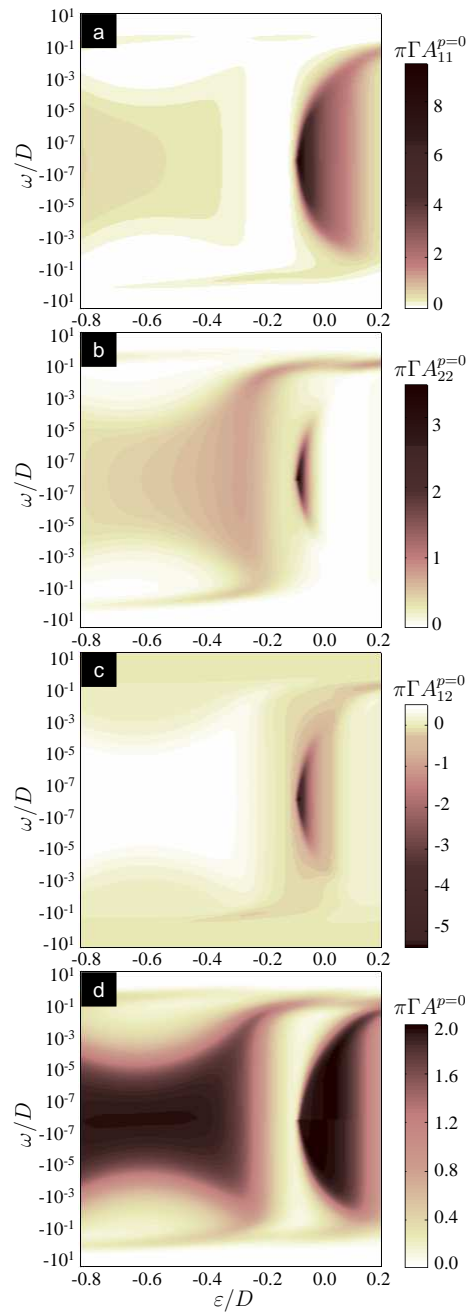}
  \caption{\label{Fig:3}
  (color online) The density plots of the normalized
  level-resolved spectral function
  $\pi\Gamma A_{jj'}^{p=0} = \pi\sum_\s \Gamma_\s A_{jj'\s}^{p=0}$ (a)-(c)
  and the full spectral function
  $\pi\Gamma A^{p=0} = \sum_{jj'}\pi\Gamma
  A_{jj'}^{p=0}$ (d) in the case when the dot is
  attached to a nonmagnetic electrode ($p=0$).
  The parameters are the same as in Fig.~\ref{Fig:2} with $\Gamma=0.04$.
  Note the logarithmic scale on the frequency axis.}
\end{figure}

Note that the abrupt jump of the dot polarization is due to the
fact that the underscreened Kondo model is extremely susceptible
to even a tiny magnetic field. The reason is that the ground state
of the system consists of a Fermi liquid and a decoupled residual
spin $S=\frac{1}{2}$ and that residual spin at zero temperature
can be polarized by any infinitesimal magnetic field. In the case
of a weakly coupled second lead there is a second stage of the
Kondo screening at $T_K'$ when the remaining spin degree of
freedom is quenched by the weakly coupled lead's
electrons.~\cite{PosazhennikovaPRB07} In that case the sudden step
in the polarization turns into a very steep crossover with a width
of $\sim \max(T_K',T)$, where $T$ is the experimental temperature.

\begin{figure}[t!]
  \includegraphics[width=1\columnwidth,clip]{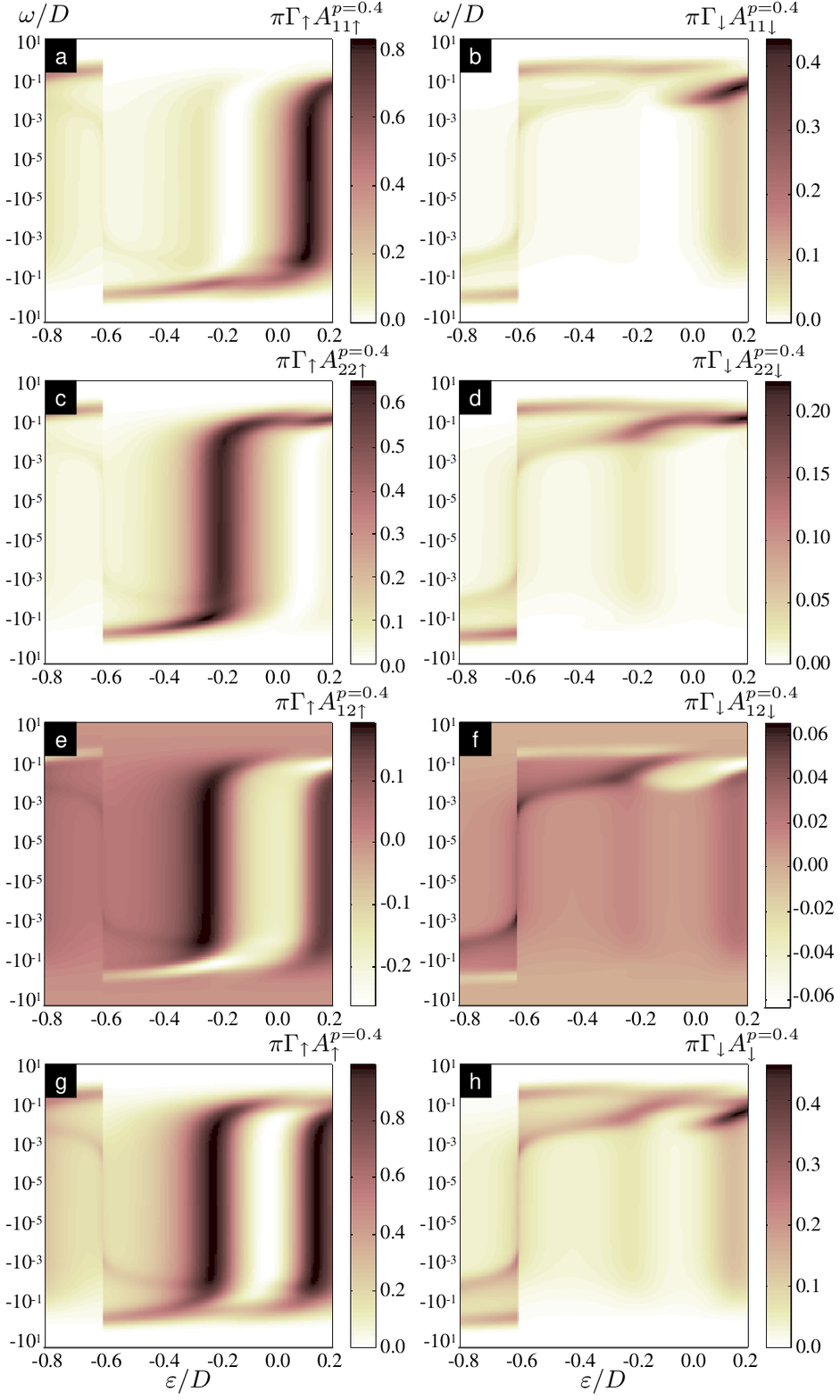}
  \includegraphics[width=0.65\columnwidth,clip]{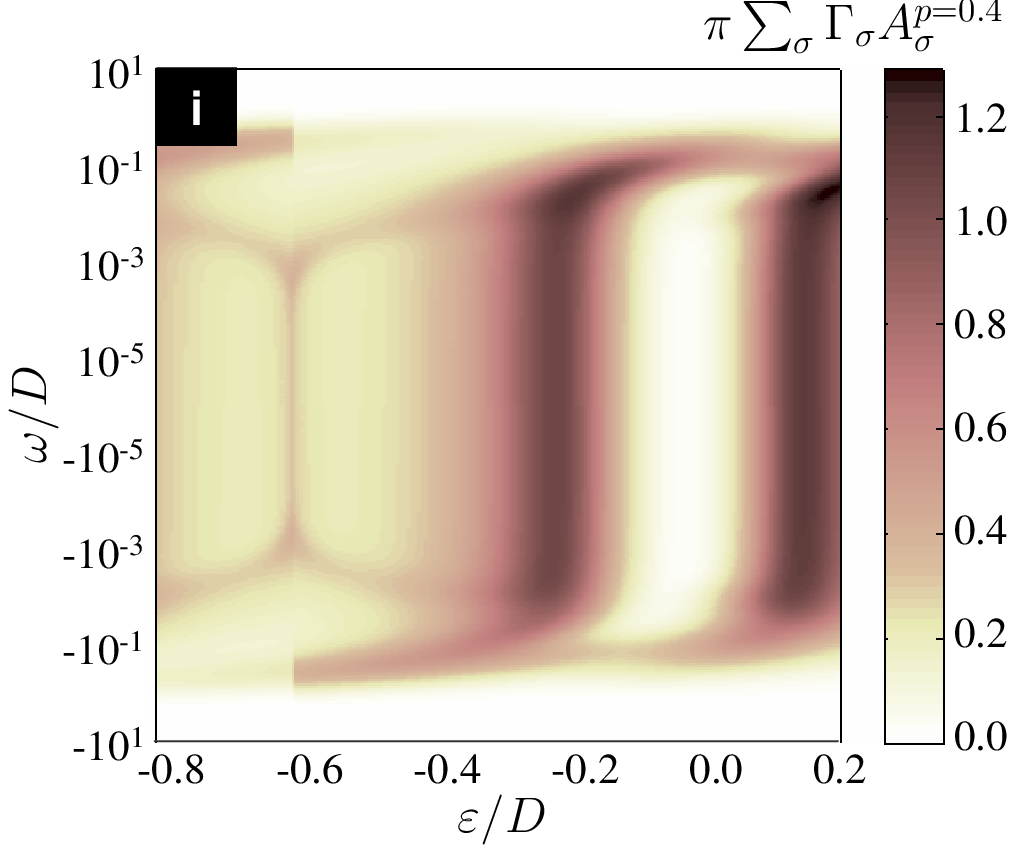}
  \caption{\label{Fig:4}
  (color online) The density plots of the spin-dependent level-resolved
  spectral function $\pi\Gamma_\s A_{jj'\s}^{p=0.4}$ (a)-(f),
  the spin-dependent full spectral function
  $\pi\Gamma_\s A_{\s}^{p=0.4} = \sum_{jj'}\pi\Gamma A_{jj'\sigma}^{p=0.4}$ (g) and (h),
  and the full spectral function
  $\pi\sum_{\s}\Gamma_\s A_{\s}^{p=0.4}$ (i)
  in the case of ferromagnetic lead with $p=0.4$.
  The parameters are the same as in Fig.~\ref{Fig:2} with $\Gamma=0.04$.
  Note the logarithmic scale on the frequency axis.}
\end{figure}

Since the dot spin polarization is rather difficult to detect
experimentally, we have computed the level-resolved single
particle spectral density $A_{jj'\s}(\omega) = -\frac{1}{2\pi}{\rm
Im}\left[G^R_{jj'\s}(\omega)+G^R_{j'j\s}(\omega)\right]$, where
$G_{jj'\s}^R(\omega)$ denotes the corresponding retarded Green's
function. This quantity contains the information about the
transport properties of the dot. Given that the dot is strongly
coupled to one of the leads, the conductivity through the setup at
voltage bias $eV$ is given by $dI/dV \sim \frac{e^2}{h} \pi
\sum_{jj'\s} \Gamma_\s A_{jj'\s} (\omega=eV)$.

The results for the level-resolved normalized spectral functions
$\pi\Gamma A_{jj'}^{p=0} = \pi \sum_\s \Gamma_\s A_{jj'\s}^{p=0}$
in the case of nonmagnetic lead ($p=0$) are shown in
Fig.~\ref{Fig:3}. The spectral functions are plotted as a function
of energy $\omega$ and average level position $\e$. To resolved
the low-energy behavior of spectral functions logarithmic scale
for $\omega$ is used. Furthermore, only the Coulomb blockade
valleys with $N=1$ and $N=2$ electrons in the dot are shown. The
behavior of spectral functions for blockade valley with $N=3$
electrons is similar to that with a single electron due to the
particle-hole symmetry.

First of all, we note that the two Coulomb blockade regimes with
$N=1$ and $N=2$ are characterized by fundamentally different fixed
points. In the $N=1$ regime ($-0.2<\varepsilon<0.2$) the effective
low-energy model is a Fermi liquid with fully screened spin
$S=\frac{1}{2}$ ($S=\frac{1}{2}$ Kondo model). On the other hand,
in the two-electron regime ($-1<\varepsilon<-0.2$) the effective
model is a singular Fermi liquid,~\cite{MehtaPRB05} i.e. a normal
Fermi liquid plus a free spin-$\frac{1}{2}$ with weak residual
ferromagnetic coupling to the conduction channel. These two
distinct ground states of the system are separated by a quantum
phase transition,~\cite{PustilnikPRB06,Logan_arXiv09} which occurs
when sweeping the gate voltage through the boundary of $N=1$
($N=3$) and $N=2$ Coulomb valleys. The quantum phase transition is
of Kosterlitz-Thouless type,~\cite{Logan_arXiv09} with
exponentially decreasing Kondo temperature when approaching the
transition from the normal Fermi liquid side. The different
behavior associated with two distinct fixed points can be observed
in the dependence of the level-resolved spectral functions. In the
$N=1$ Coulomb valley the diagonal elements of the spectral
function exceed the unitary value of $\pi\Gamma$, while the
off-diagonal elements become negative and their absolute value is
also larger than $\pi\Gamma$. On the other hand, in the $N=2$
Coulomb blockade valley, the diagonal and off-diagonal elements of
$A^{p=0}$ are rather positive and always smaller than $\pi\Gamma$.
In consequence, the full spectral function is properly normalized
$A^{p=0} = A^{p=0}_{11} + A^{p=0}_{22} + 2 A^{p=0}_{12} = 2 \pi
\Gamma$ for $\omega = 0$, see Fig.~\ref{Fig:3}(d). Furthermore,
the full spectral function clearly exhibits the resonance
corresponding to the underscreened Kondo effect in the Coulomb
blockade valley $N=2$ ($\varepsilon<-0.2$), while for $N=1$
($\e>-0.2$) the $S=\frac{1}{2}$ Kondo effect develops.

The spin and level-resolved normalized spectral functions
$\pi\Gamma A_{jj'\s}^{p=0.4}$ in the case of ferromagnetic lead
with $p=0.4$ are displayed in Fig.~\ref{Fig:4}. Now the spectral
function is different for each spin component due to the
spin-dependence of the coupling parameters. Furthermore, when the
lead is ferromagnetic, both Kondo resonances become completely
suppressed. This is directly associated with spin-dependent level
renormalization, which lifts the spin degeneracy in the dot,
suppressing the spin-flip cotunneling processes driving the Kondo
effect. In other words, the ferromagnetic lead exerts an effective
exchange field on the dot, which destroys the Kondo effect. There
is only a very narrow resonance at $\varepsilon=-0.6$ when the
spin splitting of the dot levels generated by the spin imbalance
of the conduction electrons vanishes, which happens exactly in the
middle of the triplet valley, see also Fig.~\ref{Fig:2}(b).

\begin{figure}[t]
  \includegraphics[height=13cm,clip]{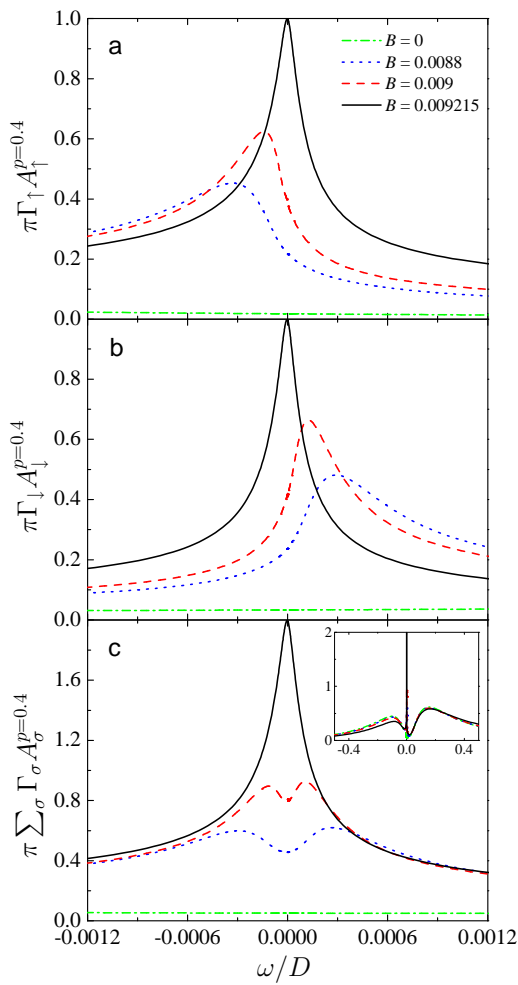}
  \caption{\label{Fig:5}
  (color online) The spin-up (a), spin-down (b) and the full (c) spectral function
  in the $S=\frac{1}{2}$ Kondo regime for $\varepsilon=0$
  in the presence of external
  magnetic field $B$, as indicated in the figure.
  The parameters are the same as in Fig.~\ref{Fig:2}
  with $\Gamma=0.04$ and $p=0.4$. The inset in (c) shows
  the zoom-out full spectral function.
  The Kondo effect is restored when $B=0.009215$.}
\end{figure}

\begin{figure}[t]
  \includegraphics[height=13cm,clip]{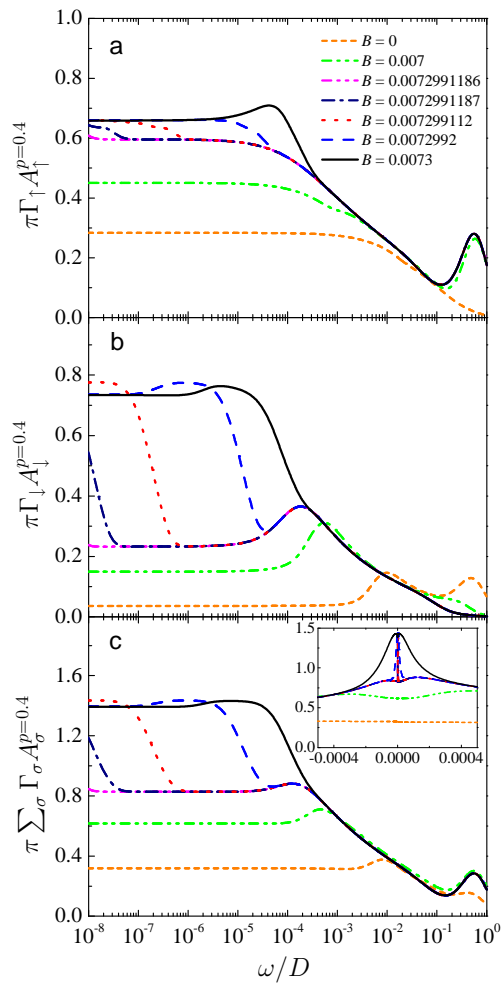}
  \caption{\label{Fig:6}
  (color online) The spin-up (a), spin-down (b) and the full (c) spectral function
  in the underscreened Kondo regime for $\varepsilon=-0.4$
  in the presence of external magnetic field $B$, as indicated in the figure.
  The parameters are the same as in Fig.~\ref{Fig:2}
  with $\Gamma=0.04$ and $p=0.4$.
  The inset in (c) displays the dependence of the spectral function
  associated with the Kondo peak.}
\end{figure}

As mentioned above, the suppression of the Kondo peaks is
associated with the exchange field which develops in the presence
of spin-dependent couplings. However, as shown in the case of
single level quantum dots,~\cite{martinek_PRB05,sindelPRB07} one
may try to compensate for the presence of the exchange field by
applying properly tuned external magnetic field. In
Fig.~\ref{Fig:5} we show the spin-resolved spectral functions
calculated in the regime of $S=\frac{1}{2}$ Kondo effect for
$\varepsilon=0$ for different values of magnetic field $B$. It can
be clearly seen that by tuning the magnetic field it is possible
to fully restore the Kondo resonance at $\omega=0$. This happens
when $B=B_{\rm c}=0.009215$, see Fig.~\ref{Fig:5}, with $B_{\rm
c}$ being the compensating field, i.e. a field at which the
previously-split dot level becomes degenerate again.

The situation, however, becomes more complicated for
$\varepsilon=-0.4$, which corresponds to the Coulomb valley with
$N=2$ and $S=1$, when the dot is described by the underscreened
Kondo model. The respective spin-resolved spectral functions are
shown in Fig.~\ref{Fig:6} using the logarithmic scale to emphasize
the distinct dependence on magnetic field. Because the dot is
coupled only to one conduction channel, only a half of the dot's
spin can be screened by conduction electrons and the ground state
consists of a Fermi liquid and a decoupled spin $S=\frac{1}{2}$.
Consequently, at $T=0$ an infinitesimally small magnetic field is
enough to polarize the residual spin-a-half in the dot. This leads
to an extremely high sensitivity of transport properties with
respect to magnetic field, as can be seen in Fig.~\ref{Fig:5}. In
consequence, in order to compensate for the presence of exchange
field in the case of underscreened Kondo model, one needs to
perform a fine-tuning of magnetic field. Furthermore, it turns out
that although it is possible to restore the resonance peak at
$\omega=0$, the full restoration of the underscreened Kondo effect
is not possible, as the height of the peak is below the unitary
limit.

\section{Conclusions}

In this paper we have analyzed the equilibrium transport
properties of a two-level quantum dot asymmetrically coupled to
ferromagnetic leads. We have shown that by tuning the position of
the dot levels, the ground state of the system changes from a
Fermi liquid in the $S=\frac{1}{2}$ Coulomb valley into a Fermi
liquid plus residual spin-a-half in the $S=1$ Coulomb valley, the
latter being the example of underscreened Kondo effect. The
boundary between these two regime is a quantum phase transition.
For finite spin polarization of the leads, $p\neq 0$, the Kondo
phenomenon becomes suppressed due to an effective exchange field
originating from the presence of ferromagnetic leads, and so is
the critical behavior.

In the transport regime where the system exhibits underscreened
Kondo effect and for $p\neq 0$, we have found that the
polarization of the dot abruptly changes sign in the middle of the
$S=1$ Coulomb blockade region. This is associated with virtual
tunneling processes (either electron-like or hole-like) that give
the dominant contribution to the renormalization of the dot
levels.

Furthermore, we have also analyzed the effect of an external
magnetic field $B$ applied to the dot. It has been shown that an
appropriately tuned $B$ can restore the Kondo effect in the case
of the $S=\frac{1}{2}$ Coulomb valley. The underscreened Kondo
effect, on the other hand, exhibits an extremely high sensitivity
towards even a tiny change in magnetic field due to its particular
ground state. Consequently, an infinitesimally small magnetic
field is enough to polarize the residual spin-a-half in the dot,
which definitely hinders a full restoration of the underscreened
Kondo effect by tuning the magnetic field.


\begin{acknowledgments}

We acknowledge fruitful discussions with J. von Delft and R.
\v{Z}itko. The authors acknowledge support from the Alexander von
Humboldt Foundation. I.W. acknowledges support from the Foundation
for Polish Science and funds of the Polish Ministry of Science and
Higher Education as research projects for years 2006-2009 and
2008-2010. Financial support by the Excellence Cluster
"Nanosystems Initiative Munich (NIM)" is gratefully acknowledged.
L.B. acknowledges the support from Hungarian Grants OTKA through
projects K73361 and NNF78842.

\end{acknowledgments}


\end{document}